\documentclass[12pt]{iopart}
\usepackage{latexsym}
\newcommand{\From}{From }       
\newcommand{\E}{{\mathcal E}}
\renewcommand{\Re}{\mbox{Re }}

\newcommand{\I}{{\rm i}}
\renewcommand{\a}{\alpha}
\renewcommand{\b}{\beta}
\newcounter{subeq}
\newenvironment{abc}
{\setcounter{subeq}{\value{equation}} \setcounter{equation}{0}
    
          \stepcounter{subeq}}
{
    \setcounter{equation}{\value{subeq}}
    \setcounter{subeq}{0}}
\begin{document}

\title[Equatorial symmetry/antisymmetry of stationary axisymmetric
    spacetimes. II]
    {Equatorial symmetry/antisymmetry of stationary axisymmetric
    electrovac spacetimes. II}
\author{Frederick J.\ Ernst\footnote[1]{e-mail: scientia@localnet.com},
    Vladimir S.\ Manko\footnote[2]{e-mail: vsmanko@fis.cinvestav.mx}
    and Eduardo Ruiz\footnote[3]{e-mail: eruiz@usal.es}}
\address{$\dag$ FJE Enterprises, 511 CR 59, Potsdam, NY 13676, USA \\
         $\ddag$ Depto.\ de F\'{\i}sica, Centro de Investigaci\'{o}n
     y de Estudios Avanzados del I.P.N., \\
     \hspace{9em} A.P.\ 14-740, 07000 M\'{e}xico, D.F., Mexico \\
     $\S$ Depto.\ de F\'{\i}sica Fundamental, Universidad de Salamanca,
     37008 Salamanca, Spain}

\begin{abstract}
In this second paper devoted to the equatorial symmetry/antisymmetry
of stationary axisymmetric electrovac spacetimes we show how two
theorems proved in our previous paper (Ernst, Manko and Ruiz 2006
{\it Class. Quantum Grav.} {\bf 23} 4945) can be utilized to construct
exact solutions that are equatorially symmetric or antisymmetric.
\end{abstract}

\section{Introduction}

Theorems which allow one to single out ``equatorially symmetric''
or ``equatorially antisymmetric'' spacetimes from the totality of
stationary axisymmetric electrovac solutions of the Einstein--Maxwell
equations were recently published \cite{EMR} by Ernst, Manko and Ruiz.
Specifically, in terms of the complex Ernst potentials \cite{Ern1}
$\E$ and $\Phi$, equatorially symmetric spacetimes satisfy the conditions
\begin{equation}
\E(-z,\rho) = \E(z,\rho)^{*},\quad\Phi(-z,\rho) =
e^{2\I\delta}\Phi(z,\rho)^{*}, \quad \delta={\rm const},
\label{cond1_sym}
\end{equation}
while equatorially antisymmetric spacetimes satisfy the conditions
\begin{equation}
\E(-z,\rho) = \E(z,\rho),\quad \Phi(-z,\rho) = \pm\Phi(z,\rho).
\label{cond1_anti}
\end{equation}
Alternatively, in terms of the axis data $e_{+}(z)$ and
$f_{+}(z)$ that comprise the values of the potentials $\E$ and
$\Phi$ on the upper part of the symmetry axis, analytically extended
to negative values of $z$, asymptotically flat stationary axisymmetric
solutions with equatorial symmetry were shown to satisfy the relations
\begin{equation}
e_{+}(z)[e_{+}(-z)]^{*}=1, \quad f_{+}(z) = -
e^{2\I\delta}[f_{+}(-z)]^{*}e_{+}(z), \label{cond2_sym}
\end{equation}
while asymptotically NUT equatorially antisymmetric solutions were
shown to satisfy the relations
\begin{equation}
e_{+}(z)e_{+}(-z)=1, \quad f_{+}(z) = \mp f_{+}(-z)e_{+}(z).
\label{cond2_anti}
\end{equation}

\begin{abc}
A large number of astrophysically relevant electrovac solutions
were considered in the papers \cite{Ern2,Ern3} and \cite{RMM,MRu1};
these correspond to rational axis data, which for some sufficiently
large value of $N$ can be expressed in the form
\begin{eqnarray}
e_{+}(z)=\frac{z^N+\sum_{k=1}^N a_k z^{N-k}} {z^N+\sum_{k=1}^N b_k
z^{N-k}} \equiv \frac{P(z)}{R(z)}, \label{data_e}\\
f_{+}(z)=\frac{\sum_{k=1}^N c_k z^{N-k}} {z^N+\sum_{k=1}^N b_k
z^{N-k}} \equiv \frac{Q(z)}{R(z)}, \label{data_f}
\end{eqnarray}
where $a_k$, $b_k$, $c_k$ are arbitrary complex parameters.  By
systematically exploiting the definition (\ref{cond1_sym}) and the
condition (\ref{cond2_sym}), we shall in this paper construct new
equatorially symmetric solutions corresponding to such rational axis
data.  Similarly, using (\ref{cond1_anti}) and (\ref{cond2_anti}),
we shall construct new equatorially antisymmetric solutions.
Finally, we shall discuss the prospects for constructing all
equatorially symmetric or antisymmetric solutions that correspond
to data (\ref{data_e}) and (\ref{data_f}) with arbitrarily large
values of $N$.  In particular, we shall show how this can be
accomplished for vacuum fields.
\end{abc}

\section{Axis data of equatorially symmetric/antisymmetric fields}

To single out equatorially symmetric axis data, we first
substitute $e_{+}(z)$ from (\ref{data_e}) into the first
condition from (\ref{cond2_sym}), whence we obtain
\begin{equation}
P(z)=(-1)^N R^{*}(-z). \label{PR_sym}
\end{equation}
\From this we obtain the first restriction upon the parameters; namely,
\begin{equation}
a_k=(-1)^k b_k^{*}, \quad k=1,\ldots,N. \label{ab_sym}
\end{equation}
The substitution of $e_{+}(z)$ and $f_{+}(z)$ from (\ref{data_e})
and (\ref{data_f}) into the second condition from (\ref{cond2_sym})
then leads, after taking account of (\ref{PR_sym}), to
\begin{equation}
Q(z)+{\rm e}^{2\I\delta}(-1)^N Q^{*}(-z)=0, \label{Q_sym}
\end{equation}
which imposes a second restriction upon the parameters; namely,
\begin{equation}
c_k+{\rm e}^{2\I\delta}(-1)^k c_k^{*}=0, \quad k=1,\ldots,N.
\label{c_sym}
\end{equation}

Turning now to the equatorially antisymmetric case, we first get from
(\ref{data_e}), (\ref{data_f}) and (\ref{cond2_anti}) the relations
\begin{equation}
P(z)=(-1)^N R(-z) \quad \mbox{and} \quad Q(z)\mp(-1)^{N+1}
Q(-z)=0, \label{PRQ_anti}
\end{equation}
and then obtain from (\ref{PRQ_anti}) the following restrictions upon
the parameters:
\begin{equation}
a_k=(-1)^k b_k, \quad c_k\mp(-1)^{k+1} c_k=0, \quad k=1,\ldots,N.
\label{abc_anti}
\end{equation}

Formulae (\ref{ab_sym}), (\ref{c_sym}) and (\ref{abc_anti}), which
provide the general forms of the axis data (\ref{data_e}) and
(\ref{data_f}) in the equatorially symmetric/antisymmetric cases,
will simplify the search for examples of axis data from which may be
constructed Ernst potentials $\E$ and $\Phi$ of spacetimes of
particular physical interest.

As an illustration of the use of the above relations let us take
$N=2$ in the axis data (\ref{data_e}) and (\ref{data_f}).  \From
(\ref{ab_sym}) we find that in the equatorially symmetric case
the function $e_{+}(z)$ has the form
\begin{equation}
e_{+}(z)=\frac{z^2-b_1^{*}z+b_2^{*}}
{z^2+b_1z+b_2}, \label{e2_sym}
\end{equation}
while from (\ref{c_sym}) we see that $f_{+}(z)$ has the form
\begin{equation}
f_{+}(z)=\frac{e^{\I\delta}(qz+\I c)} {z^2+b_1z+b_2}, \label{f2_sym}
\end{equation}
where $q$ and $c$ are arbitrary real constants.  On the other
hand, in the equatorially antisymmetric case, from (\ref{abc_anti})
one finds that
\begin{equation}
e_{+}(z)=\frac{z^2-b_1z+b_2} {z^2+b_1z+b_2}, \label{e2_anti}
\end{equation}
while for $f_{+}(z)$ one arrives at the following two possibilities:
\begin{equation}
f_{+}(z)=\frac{c_2} {z^2+b_1z+b_2} \label{f2_anti1}
\end{equation}
and
\begin{equation}
f_{+}(z)=\frac{c_1z} {z^2+b_1z+b_2}. \label{f2_anti2}
\end{equation}
Note that all parameters in formulae (\ref{e2_anti})--(\ref{f2_anti2})
may be assigned arbitrary complex values.

In the next section we shall determine the Ernst potentials $\E(z,\rho)$
and $\Phi(z,\rho)$ corresponding to three specifications of the axis
data; namely, that specified by (\ref{e2_sym}) and (\ref{f2_sym}), that
specified by (\ref{e2_anti}) and (\ref{f2_anti1}), and that specified
by (\ref{e2_anti}) and (\ref{f2_anti2}).

\section{$\E$ and $\Phi$ for rational axis data}

The general solution arising from the rational axis data
(\ref{data_e}) and (\ref{data_f}) is known \cite{RMM}, but the
description of the equatorially symmetric and antisymmetric
subfamilies of this solution in terms of the parameters it
contains has not yet been treated in any publication. Recall that
\begin{abc}
\begin{equation}
\E=E_+/E_-, \qquad \Phi=F/E_-,
\end{equation}
where
\begin{eqnarray}
&&E_\pm=\left|\begin{array}{cccc} 1 & 1 & \ldots & 1 \\ \pm 1 &
{\displaystyle \frac{r_1}{\a_1-\b_1}} & \ldots & {\displaystyle
\frac{r_{2N}}{\a_{2N}-\b_1}}\\ \vdots & \vdots & \ddots &
\vdots\\ \pm 1 & {\displaystyle \frac{r_1}{\a_1-\b_N}} & \ldots
& {\displaystyle \frac{r_{2N}}{\a_{2N}-\b_N}} \vspace{0.25cm}\\
0 & {\displaystyle \frac{h_1(\a_1)}{\a_1-\b_1^{*}}} & \ldots &
{\displaystyle \frac{h_1(\a_{2N})}{\a_{2N}-\b_1^{*}}}\\ \vdots
& \vdots & \ddots & \vdots\\ 0 & {\displaystyle
\frac{h_N(\a_1)}{\a_1-\b_N^{*}}} & \ldots & {\displaystyle
\frac{h_N(\a_{2N})}{\a_{2N}-\b_N^{*}}}\\
\end{array}\right|, \label{Edets}
\end{eqnarray}
\begin{eqnarray}
F=\left|\begin{array}{cccc} 0 & f(\a_1) & \ldots &
f(\a_{2N})
\\ -1 & {\displaystyle \frac{r_1}{\a_1-\b_1}} & \ldots &
{\displaystyle \frac{r_{2N}}{\a_{2N}-\b_1}}\\ \vdots & \vdots &
\ddots & \vdots\\ -1 & {\displaystyle \frac{r_1}{\a_1-\b_N}} &
\ldots & {\displaystyle \frac{r_{2N}}{\a_{2N}-\b_N}}
\vspace{0.25cm}\\ 0 & {\displaystyle
\frac{h_1(\a_1)}{\a_1-\b_1^{*}}} & \ldots & {\displaystyle
\frac{h_1(\a_{2N})}{\a_{2N}-\b_1^{*}}}\\ \vdots & \vdots &
\ddots & \vdots\\ 0 & {\displaystyle
\frac{h_N(\a_1)}{\a_1-\b_N^{*}}} & \ldots & {\displaystyle
\frac{h_N(\a_{2N})}{\a_{2N}-\b_N^{*}}}\\
\end{array}\right|, \label{Fdet}
\end{eqnarray}
and
\begin{eqnarray}
r_n=\sqrt{\rho^2+(z-\a_n)^2}, \\
h_l(\a_n)=e_l^{*}+2f_l^{*}f(\a_n), \quad
f(\a_n)=\sum\limits_{l=1}^N\frac{f_l}{\a_n-\b_l}, \\
e_l=\frac{2\prod_{n=1}^{2N}(\b_l-\a_n)} {\prod_{k\ne l}^{N}
(\b_l-\b_k)\prod_{k=1}^{N}(\b_l-\b_k^{*})}
-2\sum\limits_{k=1}^N\frac{f_l f_k^{*}}{\b_l-\b_k^{*}}.
\end{eqnarray}
Here $E_\pm$ and $F$ are determinants of order $2N+1$, the arbitrary
parameters are $\b_l$, $\a_n$, $f_l$, and all other constant objects
are expressible in terms of these parameters.
\end{abc}

The constants $\b_l$ can take on arbitrary complex values, while the
$\a_n$ take on arbitrary real values or occur in complex conjugate
pairs.  The parameters $f_l$ are arbitrary complex constants, and
together with the $\b_l$ they enter the axis data
\begin{equation}
e_+(z)=1+\sum\limits_{l=1}^N\frac{e_l}{z-\b_l}, \quad
f_+(z)=\sum\limits_{l=1}^N\frac{f_l}{z-\b_l}, \label{data2_2N}
\end{equation}
which is a different but mathematically equivalent representation of the
data (\ref{data_e}) and (\ref{data_f}).  The two sets of the parameters,
$\{\b_l,\a_n,f_l\}$ and $\{\b_l,e_l,f_l\}$, are mathematically
equivalent too, and they are related to each other via the algebraic
equation
\begin{equation}
S(z)\equiv e_+(z)+e_{+}^{*}(z)+2f_+(z)f_{+}^{*}(z)=0, \label{alg_eq2}
\end{equation}
which is satisfied identically by the parameters $\a_n$; i.e.,
$S(\a_n)=0$, $n=1,\ldots,2N$.

\subsection{Axis data (\ref{e2_sym}) and (\ref{f2_sym})}

The axis data (\ref{e2_sym}) and (\ref{f2_sym}) involves (in addition
to the real constant $\delta$ which we shall suppress) two complex
parameters $\{b_1,b_2\}$ and two real parameters $\{q,c\}$.  We shall
express the two complex parameters in terms of four real parameters
$\{m,a,b,k\}$ that are of more direct physical significance.  The axis
data then assume the form
\begin{abc}
\begin{eqnarray}
e_+(z)=\frac{(z-m-\I a)(z+\I b)-k} {(z+m-\I a)(z+\I b)-k},
\label{data1_ex1}\\
\quad f_+(z)=\frac{qz + \I c} {(z+m-\I a)(z+\I b)-k}.
\label{data1_ex2}
\end{eqnarray}
The solution corresponding to such axis data should be useful for
modelling the exterior field of a neutron star, for the mass--quadrupole
and angular--momentum--octupole moments it possesses are arbitrary and
independent.  This is expected to result in a generalization of two
exact solutions \cite{MMRSZ,MMS} which were recently shown to be
useful in this way \cite{SSu,SCa,BSt}.  Incidentally, there were
various typographical errors in \cite{MMR} that will be corrected
in the present paper.
\end{abc}

The first step in obtaining the potentials $\E$ and $\Phi$ from
the axis values (\ref{data1_ex1}) and (\ref{data1_ex2}) consists in
getting the corresponding expressions for $\a_n$ in terms of the
parameters $m$, $a$, $b$, $k$, $q$, $c$.  The substitution of
(\ref{data1_ex1}) and (\ref{data1_ex2}) into (\ref{alg_eq2}) yields
the biquadratic equation
\begin{equation} z^4-(m^2-a^2-b^2-q^2+2k)z^2+(k-ab)^2-b^2m^2+c^2=0,
\label{biquad1}
\end{equation}
whence we get
\begin{equation}
\a_1=-\a_4=\frac{1}{2}(\kappa_++\kappa_-), \quad
\a_2=-\a_3=\frac{1}{2}(\kappa_+-\kappa_-), \label{roots1}\\
\end{equation}
where
\begin{equation}
\kappa_\pm=\sqrt{m^2-a^2-b^2-q^2+2k\pm 2d}, \quad
d=\sqrt{(k-ab)^2-b^2m^2+c^2},
\end{equation}
thus also identifying four functions $r_n$:
\begin{equation}
r_1=R_-, \quad r_2=r_-, \quad r_3=r_+, \quad r_4=R_+, \label{rn1} \\
\end{equation}
where
\begin{equation}
R_\pm=\sqrt{\rho^2+[z\pm{\textstyle\frac{1}{2}}(\kappa_++\kappa_-)]^2},
\quad r_\pm=\sqrt{\rho^2+
[z\pm{\textstyle\frac{1}{2}}(\kappa_+-\kappa_-)]^2}.
\end{equation}

The next step is obtaining the expressions for $\b_l$, $e_l$
and $f_l$. It is easy to see that the axis data (\ref{data1_ex1})
and (\ref{data1_ex2}) can be rewritten in the form
\begin{abc}
\begin{equation}
e_+(z)=1+\frac{e_1}{z-\b_1}+\frac{e_2}{z-\b_2}, \quad
f_+(z)=\frac{f_1}{z-\b_1}+\frac{f_2}{z-\b_2}
\label{data2_ex1}
\end{equation}
where
\begin{eqnarray}
\b_1=-\frac{1}{2}[m-\I(a-b)+\sqrt{D}], \quad
\b_2=-\frac{1}{2}[m-\I(a-b)-\sqrt{D}], \quad \\
e_1=\frac{2m(\b_1+\I b)}{\b_2-\b_1}, \quad
e_2=\frac{2m(\b_2+\I b)}{\b_1-\b_2}, \\
f_1=\frac{q\b_1+\I c}{\b_1-\b_2}, \quad
f_2=\frac{q\b_2+\I c}{\b_2-\b_1}, \quad
D=[m-\I(a+b)]^2+4k, \label{data2_ex2}
\end{eqnarray}
so that now all the quantities entering the determinants (\ref{Edets})
and (\ref{Fdet}) are defined in terms of the parameters of the axis
data (\ref{data1_ex1}) and (\ref{data1_ex2}) when we take into account
that the $f(\a_n)$ are obtainable from $f_+(z)$ by simply changing $z$
to $\a_n$,
\end{abc}
\begin{equation}
f(\a_n)=\frac{q\a_n + \I c} {(\a_n+m-\I a)(\a_n+\I b)-k},
\label{fan1}
\end{equation}
while $h_l(\a_n)$ are combinations of $e_l$, $f_l$ and $f(\a_n)$.

\begin{abc}
The final step is the expansion of the determinants (\ref{Edets})
and (\ref{Fdet}), which in the case $N=2$ have the order 5, followed
by the substitution of $e_l$, $f_l$, $\b_l$, $f(\a_n)$ and
$h_l(\a_n)$ as given by the formulae (\ref{roots1}), (\ref{rn1}),
(\ref{data2_ex1})--(\ref{data2_ex2}) and (\ref{fan1}).  The
evaluation and simplification of the resulting expressions can be
carried out with the aid of a symbolic computer programme.  The final
formulae for $\E$ and $\Phi$ are\footnote{The expression for $A$
in \cite{MMR} (see p.~3068) was published with some misprints. However,
all the rest of the formulae determining the metric functions (3.11)
of \cite{MMR} are free of misprints, albeit there is a missing right
square bracket in the second line of the function $D$ just after the
first curly bracket.}
\begin{equation}
\E=\frac{A-B}{A+B}, \quad \Phi=\frac{C}{A+B},
\end{equation}
where
\begin{eqnarray}
A&=&d[m^2-q^2-(a-b)^2][\kappa_+^2(R_+r_-+R_-r_+)+
\kappa_-^2(R_+r_++R_-r_-)] \nonumber\\
&+&[(a-b)^2(ab-k)+(a-b)(cq-bm^2)+m^2b^2-c^2] \nonumber\\
&\times&[\kappa_+^2(R_+r_-+R_-r_+)-
\kappa_-^2(R_+r_++R_-r_-)] \nonumber\\
&+&4d[k(m^2-q^2)+(c+aq)(bq-c)](R_+R_-+r_+r_-) \nonumber\\
&+&\I\kappa_+\kappa_-\{\kappa_+[(a-b)(ab-k-d)-m^2b+cq]
(R_+r_--R_-r_+) \nonumber\\
&+&\kappa_-[(a-b)(ab-k+d)-m^2b+cq]
(R_-r_--R_+r_+)\}, \\
B&=&m\kappa_+\kappa_-\{ d[\kappa_+\kappa_-(R_++R_-+r_++r_-)
\nonumber\\ &-&(m^2-a^2+b^2-q^2)(R_++R_--r_+-r_-)] \nonumber\\
&+&\I bd[(\kappa_++\kappa_-)(R_+-R_-)
+(\kappa_+-\kappa_-)(r_--r_+)]
\nonumber\\
&+&\,\,
\I(m^2b-cq+ak-a^2b) \nonumber\\
&\times &[(\kappa_++\kappa_-)(r_+-r_-)+(\kappa_+-\kappa_-)(R_--R_+)]\}, \\
C&=&\kappa_+\kappa_-\{ d[q(m^2-(a-b)^2-q^2)-2c(a-b)]
(r_++r_--R_+-R_-) \nonumber\\
&+&dq\kappa_+\kappa_-(R_++R_-+r_++r_-)
+\I cd[(\kappa_++\kappa_-)(R_+-R_-) \nonumber\\
&+&(\kappa_+-\kappa_-)(r_--r_+)] +\I[(k-ab)(c+aq-bq)+q(m^2b-cq)]
\nonumber\\
&\times&[(\kappa_++\kappa_-)(r_+-r_-)
+(\kappa_+-\kappa_-)(R_--R_+)]\}.
\end{eqnarray}
\end{abc}

\subsection{Axis data (\ref{e2_anti}) and (\ref{f2_anti1})}

\begin{abc}
Proceeding in a similar way, the axis data
(\ref{e2_anti}) and (\ref{f2_anti1}) can be reexpressed in the form
\begin{eqnarray}
e_{+}(z) = \frac{z-k-m-\I(a+\nu)}{z-k+m-\I(a-\nu)} \cdot
\frac{z+k-m+\I(a-\nu)}{z+k+m+\I(a+\nu)}, \label{data_ex2_1}\\
f_{+}(z) = \frac{2(q+\I b)z}{[z-k+m-\I(a-\nu)][z+k+m+\I(a+\nu)]},
\label{data_ex2_2}
\end{eqnarray}
where $m$, $a$, $\nu$, $k$, $q$ and $b$ are arbitrary real
parameters.  Expressions (\ref{data_ex2_1}) and (\ref{data_ex2_2})
generalize the axis data of the Bret\'on--Manko solution \cite{BMa}
and can be interpreted as representing two identical counter--rotating
electrically and magnetically charged masses possessing a NUT
parameter, $m$, $q$ and $b$ being, respectively, the individual mass,
electric and magnetic charges, $a$ the rotational parameter, $\nu$ the
NUT parameter and $k$ the separation constant.
\end{abc}

Repeating now the steps outlined in the first example, this time
starting with the axis data (\ref{data_ex2_1}) and (\ref{data_ex2_2}),
one finally arrives at the following expressions for the potentials
$\E$ and $\Phi$:
\begin{abc}
\begin{equation}
\E=\frac{A-B}{A+B}, \quad \Phi=\frac{C}{A+B}, \label{ex2_1}
\end{equation}
where
\begin{eqnarray}
A&=&(m^2+\nu^2-q^2-b^2)\{[(m^2+k^2+\nu^2+a^2)^2-4(mk+a\nu)^2] \nonumber\\
&\times&(R_+-R_-)(r_+-r_-)-\delta\a_+\a_-(R_++R_-)(r_++r_-)\} \nonumber\\
&+&2\a_+\a_-\{[(m^2+\nu^2-q^2-b^2)(m^2-k^2+a^2-\nu^2)
+2(m\nu-ka)^2] \nonumber\\ &\times&(R_+R_-+r_+r_-)
+2\I d(m\nu-ka)(R_+R_--r_+r_-)\}, \\
B&=&4d\a_+\a_-(m+\I\nu)\{[m^2+\nu^2-q^2-b^2+\I(m\nu-ka)] \nonumber\\
&\times&(R_++R_--r_+-r_-)-d(R_++R_-+r_++r_-)\}, \\
C&=&4d\a_+\a_-(q+\I b)\{[m^2+\nu^2-q^2-b^2+\I(m\nu-ka)] \nonumber\\
&\times&(R_++R_--r_+-r_-)-d(R_++R_-+r_++r_-)\} \nonumber\\
&=&(q+\I b)B/(m+\I\nu),
\end{eqnarray}
and
\begin{eqnarray}
R_\pm&=&\sqrt{\rho^2+(z\pm\a_+)^2}, \quad
r_\pm=\sqrt{\rho^2+(z\pm\a_-)^2}, \\
\a_\pm&=&\sqrt{\delta\pm2d}, \quad
\delta=m^2+k^2+3\nu^2-a^2-2(q^2+b^2), \\
d&=&\sqrt{(m^2+\nu^2-q^2-b^2)(k^2+2\nu^2-a^2-q^2-b^2)-(m\nu-ka)^2}.
\label{ex2_2}
\end{eqnarray}
\end{abc}

It is easy to verify that the Ernst potentials given by
(\ref{ex2_1})--(\ref{ex2_2}) satisfy the conditions
\begin{equation}
\E(-z,\rho) = \E(z,\rho),\quad\Phi(-z,\rho) = \Phi(z,\rho).
\label{relEF_anti1}
\end{equation}
It should also be pointed out that the solution so defined is not
asymptotically flat due to the presence of the non--zero NUT
parameter $\nu$. When $\nu=b=0$, it reduces to the Bret\'on--Manko
electrovac solution \cite{BMa}.

\subsection{Axis data (\ref{e2_anti}) and (\ref{f2_anti2})}

\begin{abc}
We can reexpress the axis data (\ref{e2_anti}) and (\ref{f2_anti2})
in the form
\begin{eqnarray}
e_{+}(z) = \frac{z-k-m-\I(a+\nu)}{z-k+m-\I(a-\nu)} \cdot
\frac{z+k-m+\I(a-\nu)}{z+k+m+\I(a+\nu)}, \\
f_{+}(z) = \frac{2(\chi+\I c)}{[z-k+m-\I(a-\nu)][z+k+m+\I(a+\nu)]},
\end{eqnarray}
where the four parameters $m$, $a$, $\nu$ and $k$ have the same
meaning as in the previous example, while $\chi$ and $c$ describe,
respectively, the electric--dipole and magnetic--dipole moments of
the sources.
\end{abc}

\begin{abc}
The corresponding expressions for the potentials $\E$ and $\Phi$,
obtainable from the determinants (\ref{Edets}) and (\ref{Fdet}),
have the form
\begin{equation}
\E=\frac{A-B}{A+B}, \quad \Phi=\frac{C}{A+B}, \label{ex3_1}
\end{equation}
where
\begin{eqnarray}
A&=&\{\delta[(m^2+\nu^2)^2+(m\nu-ka)^2] -d^2(3m^2-k^2+\nu^2+a^2)\}
\nonumber\\ &\times&(R_+-R_-)(r_+-r_-)
-\a_+\a_-[\delta(m^2+\nu^2)-\chi^2-c^2] \nonumber\\
&\times&(R_++r_-)(r_++r_-)+2\a_+\a_-\{[(m^2+\nu^2)^2+(m\nu-ka)^2-d^2]
\nonumber\\
&\times&(R_+R_-+r_+r_-)
+2\I d(m\nu-ka)(R_+R_--r_+r_-)\}, \\
B&=&4d\a_+\a_-(m+\I\nu)\{[m^2+\nu^2+\I(m\nu-ka)] \nonumber\\
&\times&(R_++R_--r_+-r_-)-d(R_++R_-+r_++r_-)\}, \\
C&=&4d(\chi+\I c)\{[m^2+\nu^2-\I(m\nu-ka)]
[\a_-(R_--R_+)-\a_+(r_--r_+)]
\nonumber\\
&+&d[\a_-(R_--R_+)+ \a_+(r_--r_+)]\},
\end{eqnarray}
and
\begin{eqnarray}
R_\pm&=&\sqrt{\rho^2+(z\pm\a_+)^2}, \quad
r_\pm=\sqrt{\rho^2+(z\pm\a_-)^2}, \\
\a_\pm&=&\sqrt{\delta\pm2d}, \quad
\delta=m^2+k^2+3\nu^2-a^2, \label{delta}\\
d&=&\sqrt{(m^2+\nu^2)(k^2+2\nu^2-a^2)-(m\nu-ka)^2-\chi^2-c^2}.
\label{ex3_2}
\end{eqnarray}
Please note that the $\delta$ defined in (\ref{delta}) should not
be confused with the phase $\delta$ employed in the symmetric case!
\end{abc}

The Ernst potentials given by (\ref{ex3_1})--(\ref{ex3_2}) satisfy
the conditions
\begin{equation}
\E(-z,\rho) = \E(z,\rho),\quad\Phi(-z,\rho) = -\Phi(z,\rho),
\label{relEF_anti2}
\end{equation}
and the solution describes two counter--rotating masses endowed
with the NUT parameter and with both electric and magnetic dipole
moments.  When $\nu=0$, the solution is asymptotically flat.

\section{Prospects for a completely general solution}

A similar procedure to that which we employed for $N=2$ should work
also for $N=3$ and $N=4$, but the question arises whether or not one
can solve the general equatorially symmetric/antisymmetric problem
for arbitrarily large values of $N$.  At the present time, we do not
have an answer to this question, but in this section we shall show
that in the vacuum case this is indeed possible, which gives us hope
that the same will ultimately be possible in the case of non-vacuum
electrovac fields.

Instead of the parameters $\b_k$, the quantities
\begin{equation}
X_n\equiv X(\a_n)=\frac{\prod_{l=1}^{N}(\a_n-\b_l^{*})}
{\prod_{l=1}^{N}(\a_n-\b_l)}, \label{def_Xn}
\end{equation}
can be used in the vacuum solution, in which case the formulae
for $\E$ assume the form \cite{Ern2,MRu1}
\begin{abc}
\begin{equation}
\E=\frac{\Lambda+\Gamma}{\Lambda-\Gamma}, \label{E_vac1}
\end{equation}
where
\begin{eqnarray}
\Lambda=\sum_{i_1<\ldots<i_{N}}\lambda_{i_{1}\ldots
i_{N}}r_{i_{1}}\ldots r_{i_{N}}, \\
\Gamma=\sum_{i_1<\ldots<i_{N-1}}\gamma_{i_{1}\ldots i_{N-1}
}r_{i_{1}}\ldots r_{i_{N-1}},
\end{eqnarray}
and
\begin{eqnarray}
\lambda_{i_{1}\ldots i_{N}}=(-1)^{i_{1}+\ldots+i_{N}}
V(\a_{i_{1}},\ldots,\a_{i_{N}})
V(\a_{i'_{1}},\ldots,\a_{i'_{N}})\prod_{n=1}^NX_{i_{n}}, \nonumber\\
\hspace{1cm} (i'_{1}<\ldots<i'_{N}, i^{'}\ne i) \\
\gamma_{i_{1}\ldots i_{N-1}}=(-1)^{i_{1}+\ldots+i_{N-1}}
V(\a_{i_{1}},\ldots,\a_{i_{N-1}})
V(\a_{i'_{1}},\ldots,\a_{i'_{N+1}}) \prod_{n=1}^{N-1}X_{i_{n}},
\nonumber\\
\hspace{1cm} (i'_{1}<\ldots<i'_{N+1}, i^{'}\ne i) \\
V(\a_1,\ldots,\a_k)=\prod\limits_{1\le i<j\le k}(\a_i-\a_j).
\label{E_vac2}
\end{eqnarray}
\end{abc}

In order to show how the subclasses of the equatorially symmetric
or antisymmetric solutions are contained in (\ref{E_vac1})--(\ref{E_vac2})
we must find $2N$ restrictions on the parameters $\a_n$ and $X_n$ which
would be equivalent to $N$ conditions (\ref{ab_sym}) or
(\ref{abc_anti}) on the {\it complex} constants $a_k$.
The conditions on the parameters $\a_n$ can be trivially
established since these parameters determine the location of
sources on the symmetry axis. Assuming without loss of generality
the order
\begin{equation} \Re(\a_{1})\ge\Re(\a_{2})\ge\ldots\ge
\Re(\a_{2N-1})\ge\Re(\a_{2N}) \label{order}
\end{equation}
we immediately arrive at $N$ relations between the parameters
$\a_n$.  These relations are the same in the equatorially symmetric
and antisymmetric cases; namely,
\begin{equation}
\a_{1}=-\a_{2N}, \quad \a_{2}=-\a_{2N-1}, \quad
\ldots, \quad \a_{N}=-\a_{N+1}. \label{alphas1}
\end{equation}
A simpler way to write this is
\begin{equation}
\a_{k}=-\a_{2N-k+1}, \quad k=1,\ldots,N.  \label{alphas2}
\end{equation}
The above conditions determine the equatorially
symmetric/antisymmetric character of the mass distribution.

The restrictions on the quantities $X_n$ can be obtained from
the vacuum specialization of equation (19); namely
\begin{equation}
e_+(z)+e_+^{*}(z) = 0 \label{alg_eq}
\end{equation}
with the aid of the conditions
(\ref{PR_sym}) and (\ref{PRQ_anti}). In the equatorially symmetric
case the substitution of
\begin{equation}
e_+(z) = 1+\sum_{l=1}^{N}\frac{e_l}{z-\b_l} = \frac{P(z)}{R(z)}
\label{data_vac}
\end{equation}
into (\ref{alg_eq}) gives, after taking into account (\ref{PR_sym}),
\begin{equation}
\frac{P(z)}{R(z)}+\frac{P^{*}(z)}{R^{*}(z)}
=\frac{(-1)^N}{R(z)R^{*}(z)} [R^{*}(z)R^{*}(-z)+R(z)R(-z)]=0,
\end{equation}
whence, bearing in mind that $\a_n$ satisfy the above equation
identically, we get
\begin{equation}
\frac{R^{*}(\a_n)R^{*}(-\a_n)} {R(\a_n)R(-\a_n)} +1=0 \quad
\Longleftrightarrow \quad X(\a_n)X(-\a_n)=-1.  \label{X_sym1}
\end{equation}
Therefore, taking into account
(\ref{alphas2}), we arrive at the remaining $N$ restrictions on
the parameters $X_n$; namely,
\begin{equation}
X_{k}X_{2N-k+1}=-1, \quad k=1,\ldots,N.  \label{X_sym2}
\end{equation}

In the equatorially antisymmetric case we proceed in the analogous
way.  \From equation (\ref{alg_eq}), the axis data (\ref{data_vac}) and
the first condition from (\ref{PRQ_anti}), we first arrive at the
equation
\begin{equation}
\frac{(-1)^N}{R(z)R^{*}(z)} [R(z)R^{*}(-z)+R^{*}(z)R(-z)]=0,
\end{equation}
which is satisfied by all $\a_n$ identically.  Then we get
\begin{equation}
\frac{R^{*}(\a_n)R(-\a_n)} {R(\a_n)R^{*}(-\a_n)} +1=0 \quad
\Longleftrightarrow \quad X(\a_n)=-X(-\a_n), \label{X_anti1}
\end{equation}
thus arriving at the required restrictions on $X_n$ that characterize
the equatorially antisymmetric solutions; namely,
\begin{equation}
X_{k}=-X_{2N-k+1}, \quad k=1,\ldots,N.  \label{X_anti2}
\end{equation}

The above identification of the equatorially
symmetric/antisymmetric vacuum soliton solutions, carried out
directly in terms of the parameters entering the expression of the
Ernst potential defined by formulae (\ref{E_vac1})--(\ref{E_vac2}),
permits the consideration of particular configurations with additional
equatorial symmetries independently of the corresponding axis
data, which sometimes turns out to be very advantageous during the
treatment of specific problems.  In the papers \cite{MRM,HMM1}, for
instance, the $N=3$ and $N=4$ specializations of the relations
(\ref{alphas2}) and (\ref{X_sym2}) were successfully employed for
the analysis of equilibrium states in the triple- and
quadruple--Kerr solutions.

\section{Concluding remarks}

In this paper we have emphasized the equatorially symmetric and
antisymmetric electrovac solutions constructed from rational axis
data. Such solutions have direct relevance to multi--black--hole
spacetimes and to the fields of astrophysical objects.  When an
electrovac solution does not arise from rational axis data, the
fundamental theorem proved in our previous paper can be employed
to establish whether or not the solution in question is
equatorially symmetric/antisymmetric.  For example, from the
conditions (\ref{cond1_sym}) one can very easily establish the
equatorial symmetry of the electrovac solution for a charged
rotating Curzon mass \cite{HMM2}.  We should also like to point
out that J.\ Sod--Hoffs, who is a student of V.~S.~Manko, is
currently working on the derivation of all the metrical functions
for two of our antisymmetric solutions.  He intends to submit for
publication a paper concerning his analysis of the equilibrium
conditions as well as the physical interpretation of these
solutions.

\ack This work was partially supported by Project 45946--F from
CONACyT of Mexico, Project BFM2003--02121 from MCyT and Project
SA010C05 from Junta de Castilla y Leon, Spain.


\section*{References}
\begin{thebibliography}{10}
\bibitem{EMR}
Ernst~F~J, Manko~V~S and Ruiz~E 2006 Equatorial
symmetry/antisymmetry of stationary axisymmetric electrovac
spacetimes, {\it Class.\ Quantum Grav.}\ \textbf{23} 4945
\bibitem{Ern1}
Ernst~F~J 1968 New formulation of the axislly symmetric
gravitational field problem. II, {\it Phys.\ Rev.}\ \textbf{168}
1415
\bibitem{Ern2}
Ernst~F~J 1994 Determining parameters of the Neugebauer family of
vacuum spacetimes in terms of data specified on the symmetry axis,
{\it Phys.\ Rev.\ D}\ \textbf{50} 4993
\bibitem{Ern3}
Ernst~F~J 1994 Fully electrified Neugebauer spacetimes, {\it
Phys.\ Rev.\ D}\ \textbf{50} 6179
\bibitem{RMM}
Ruiz~E, Manko~V~S and Mart\'\i n~J 1995 Extended $N$--soliton
solution of the Einstein--Maxwell equations, {\it Phys.\ Rev.\ D}\
\textbf{51} 4192
\bibitem{MRu1}
Manko~V~S and Ruiz~E 1998 Extended multi--soliton solutions of the
Einstein field equations, {\it Class.\ Quantum Grav.}\ \textbf{15}
2007
\bibitem{MMRSZ}
Manko~V~S, Mart\'\i n~J, Ruiz~E, Sibgatullin~N~R and Zaripov~M~N
1994 Mertic of a rotating, charged, magnetized, deformed mass,
{\it Phys.\ Rev.\ D}\ \textbf{49} 5144
\nonum Manko~V~S, Mart\'\i
n~J and Ruiz~E 1994 Mertic of a rotating, charged, magnetized,
deformed mass. II, {\it Phys.\ Rev.\ D}\ \textbf{49} 5150
\bibitem{MMS}
Manko~V~S, Mielke~E~W and Sanabria--G\'omez~J~D 2000 Exact
solution for the exterior field of a rotating neutron star, {\it
Phys.\ Rev.\ D}\ \textbf{61} 081501
\nonum Manko~V~S,
Sanabria--G\'omez~J~D and Manko~O~V 2000 Nine--parameter
electrovac metric involving rational functions, {\it Phys.\ Rev.\
D}\ \textbf{62} 044048
\bibitem{SSu}
Sibgatullin~N~R and Sunyaev~R~A 1998 Disk accretion in
gravitational field of a rapidly rotating neutron star with a
rotationally induced quadrupole mass moment, {\it Astron.\ Lett.}\
\textbf{24} 774
\bibitem{SCa}
Stute~M and Camenzind~M 2002 Towards a self-consistent
relativistic model of the exterior gravitational field of rapidly
rotating neutron stars, {\it MNRAS}\ \textbf{336} 831
\bibitem{BSt}
Berti~E and Stergioulas~N 2004 Approxiate matching of analytic and
numerical solutions for rapidly rotating neutron stars, {\it
MNRAS}\ \textbf{350} 1416
\bibitem{MMR}
Manko~V~S, Mart\'\i n~J and Ruiz~E 1995 Extended Six--parameter
solution of the Einstein--Maxwell equations possessing equatorial
symmetry, {\it J.\ Math.\ Phys.}\ \textbf{36} 3063
\bibitem{BMa}
Bret\'{o}n~N and Manko~V~S 1995 A binary system of `antisymmetric'
Kerr--Newman masses, {\it Class.\ Quantum Grav.}\ \textbf{12} 1969
\bibitem{MRM}
Manko~V~S, Ruiz~E and Manko~O~V 2000 Is equilibrium of aligned
Kerr black holes possible? {\it Phys.\ Rev.\ Lett.}\ \textbf{85}
5504
\nonum Manko~O~V, Manko~V~S and Ruiz~E 2002 Equilibrium of
three collinear Kerr particles, {\it Phys.\ Rev.\ D}\ \textbf{65}
084027
\bibitem{HMM1}
Hern\'andez--Pastora~J~L, Manko~O~V, Manko~V~S,
Mart\'\i n~J and Ruiz~E 2004 Equilibrium states in the
quadruple--Kerr solution, {\it Gen.\ Relat.\ Grav.}\ \textbf{36}
781
\bibitem{HMM2} Hern\'andez--Pastora~J~L, Manko~V~S and
Mart\'\i n~J 1993 Some asymptotically flat generalizations of the
Curzon metric, {\it J.\ Math.\ Phys.}\ \textbf{34} 4760

\end{thebibliography}
\end{document}